\documentclass{IEEEoj}
\usepackage{cite}
\usepackage{amsmath,amssymb,amsfonts}
\usepackage{algorithmic}
\usepackage{graphicx,color}
\usepackage{textcomp}
\def\BibTeX{{\rm B\kern-.05em{\sc i\kern-.025em b}\kern-.08em
    T\kern-.1667em\lower.7ex\hbox{E}\kern-.125emX}}
\AtBeginDocument{\definecolor{ojcolor}{cmyk}{0.93,0.59,0.15,0.02}}
\def\OJlogo{\vspace{-14pt}\includegraphics[height=28pt]{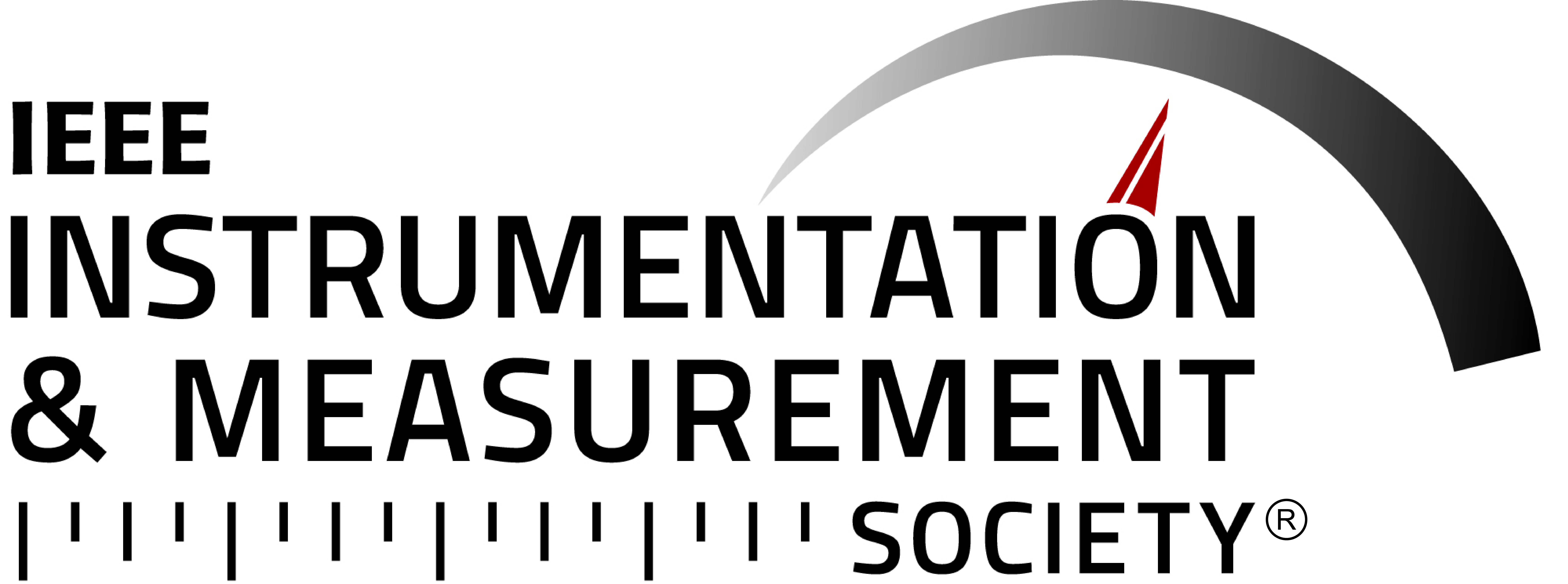}}
\begin{document}
\receiveddate{26 September, 2022}
\reviseddate{07 October, 2022}
\accepteddate{19 October, 2022}
\publisheddate{XX Month, XXXX}
\currentdate{XX Month, XXXX}
\doiinfo{OJIM.2022.1234567}

\title{When Digital Economy Meets Web3.0: Applications and Challenges}

\author{CHUAN CHEN\authorrefmark{1}, MEMBER, IEEE, LEI ZHANG\authorrefmark{1}, YIHAO LI\authorrefmark{1}, TIANCHI LIAO\authorrefmark{2}, SIRAN ZHAO\authorrefmark{1}, ZIBIN  ZHENG\authorrefmark{2}, SENIOR MEMBER, IEEE, HUAWEI HUANG\authorrefmark{2}, SENIOR MEMBER, IEEE, JIAJING WU\authorrefmark{1}, SENIOR MEMBER, IEEE}
\affil{School of Computer Science and Engineering, Sun Yat-sen University, Guangzhou, 510275, China}
\affil{School of Software Engineering, Sun Yat-sen University, Zhuhai, 519000, China}
\corresp{CORRESPONDING AUTHOR: ZIBIN ZHENG (e-mail: zhzibin@mail.sysu.edu.cn).}
\authornote{This work was supported by the National Natural Science Foundation of China (62032025, 62176269), the Guangdong Basic and Applied Basic Research Foundation (2019A1515011043), and the Technology Program of Guangzhou, China (202103050004).}

\begin{abstract}
With the continuous development of web technology, Web3.0 has attracted a considerable amount of attention due to its unique decentralized characteristics. The digital economy is an important driver of high-quality economic development and is currently in a rapid development stage. In the digital economy scenario, the centralized nature of the Internet and other characteristics usually bring about security issues such as infringement and privacy leakage. Therefore, it is necessary to investigate how to use Web3.0 technologies to solve the pain points encountered in the development of the digital economy by fully exploring the critical technologies of digital economy and Web3.0. In this paper, we discuss the aspects of Web3.0 that should be integrated with the digital economy to better find the entry point to solve the problems by examining the latest advances of Web3.0 in machine learning, finance, and data management. We hope this research will inspire those who are involved in both academia and industry, and finally help to build a favourable ecology for the digital economy.
\end{abstract}

\begin{IEEEkeywords}
    Web3.0, Digital Economy, Blockchain, Privacy Computing, DAO, Metaverse
\end{IEEEkeywords}

\maketitle
\section{INTRODUCTION}
\IEEEPARstart{T}{he} rapid growth of the digital economy in recent years has been fueled by a fresh round of technical revolution and industrial change, and the digital economy is playing an increasingly significant role in the global economy.
The digital economy is an essential driver of high-quality economic development. 
Tapscott, an American economist, originally put forth the idea of the digital economy in his book “\textit{The Digital Economy}” in 1994 \cite{bowman1996digital}.
When the Internet was still in its infancy, the primary characteristic of the digital economy at the time was the utilization of information in the network for digital flow and transmission. Tapscott predicted that the Internet will fundamentally alter both the global economy and society.
In 1998, the U.S. Department of Commerce released a report on the new digital economy and adopted the term “\textit{digital economy}” for the first time, and thus the concept of the digital economy has gradually been widely recognized by governments and scholars\cite{united1998emerging}.

With the progress of the era, the digital economy is developing rapidly and its great potential has been recognized worldwide. The digital economy is increasingly influential in people's lives, and has now become the most dynamic and innovative economic form.
Nowadays, the digital economy has been acting in various fields, gaining prominence in optimizing the economic structure and promoting industrial innovation. 
Obviously, the digital economy is an essential part of global economic development \cite{li2020should}, becoming a key role in restructuring global factor resources, reshaping the global economic structure, and changing the global competitive landscape.

Digital economy is a new form of economic operation that emerged in the late stage of industrial economy development, which utilizes information network as the major carrier and makes digital resources as production factors \cite{pozdnyakova2019internet}.
Through modern communication and Internet technologies, the digital economy makes effective usage of resources in various industries of society, thereby creating higher economic benefits than traditional industrial economy operations \cite{kagermann2015change}.
Although the development of technology has changed the operation and transaction methods of enterprises and accelerated the reliance of consumers on the Internet, the simple movement of services from “offline” to “online” has been insufficient to meet the demand for online services in the new century \cite{choi2000future}. Therefore, many companies are investing significant time and financial resources in providing virtual reality.
Thereby, in the process of digital economy development, a series of new information technologies such as cloud computing \cite{dincua2019determinants,strommen2019digital}, big data \cite{tan2017using, novikov2020data}, artificial intelligence \cite{chui2017artificial,bahtizin2019using}, have been gradually spawned. Thus, it provides technical support to further expand the scale of new industries and economies. A new economic era with the Internet as the main driving force for the economic and social development of each country has been opened.
\begin{figure*}[h]
    \centering
    \caption{The evolution progress of web technologies}
    \label{webevo}
    \includegraphics[scale=0.47]{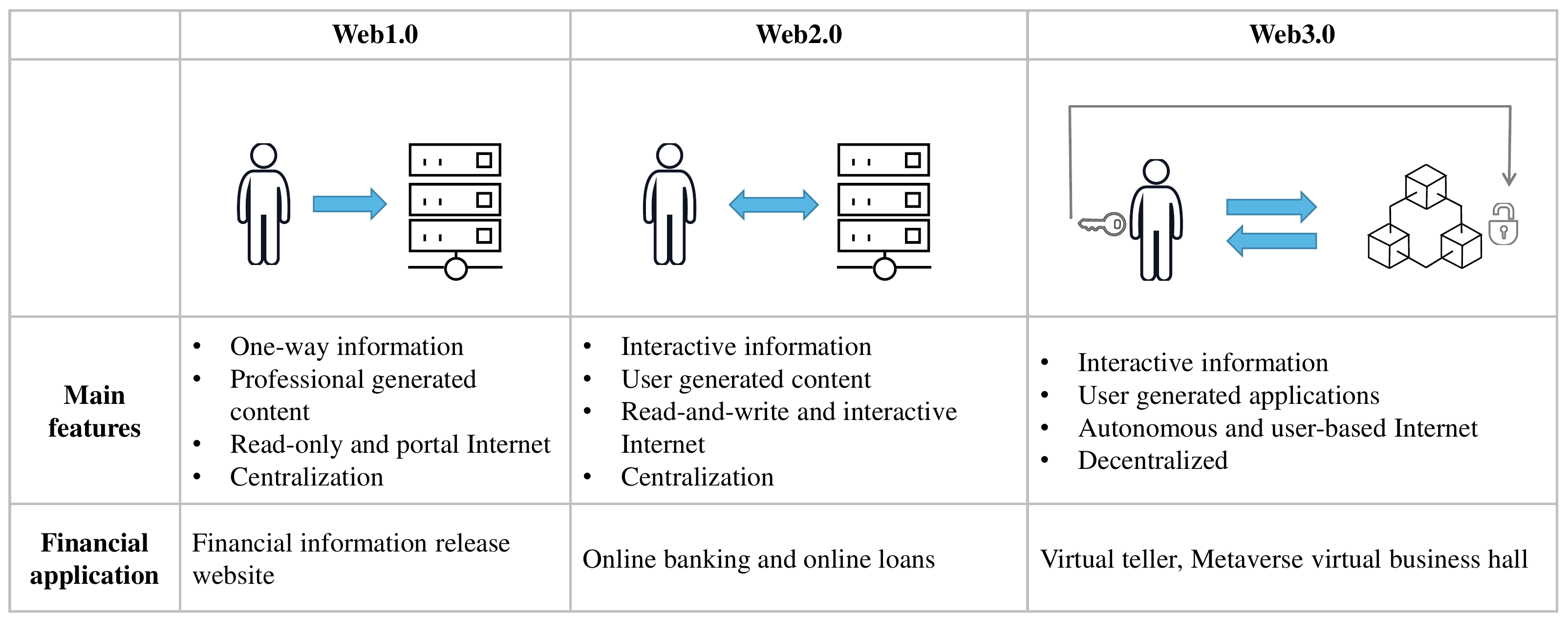}
\end{figure*}
Today, the development of information technology has allowed the digital economy to drive a comprehensive overhaul of production methods, lifestyles, and governance.
The popularity of the Internet has led to the massive rise of online economic industries such as social networking, digital art, and virtual worlds, etc., while also making security issues such as copyright infringement, and privacy breaches easier and more prevalent.
At present, unified rules for measuring the value of data have not yet been formed, the digitization of industry has not yet been completed, and digital technology has not yet been industrialized.
As we all know, each time a user interacts via the Internet, he or she sends data to the service provider who provides the server, which can easily lead to data security, privacy, and control issues \cite{acquisti2016economics,kerber2016digital}. Therefore, a “decentralized” Internet based on blockchain distributed storage technology, i.e., Web3.0, has been developed in recent years. 

Web3.0 is described as the potential next phase of the Internet, which is a “decentralized” Internet running on top of blockchain-related technologies.
In Web3.0, data presents a distributed storage structure, so that there will be no central node for data management, which significantly reduces the service cost of managing data \cite{cui2012research}. Therefore, the digital economy platform established based on Web3.0 technology is the new trend of the current development.

Since the creation of the World Wide Web in 1989 \cite{gillies2000web}, the Internet has experienced the Web1.0 era with TCP/IP and other open protocols as the underlying technologies.
Web1.0 is mainly characterized by the user's access to information provided and understood in a single direction. Users can usually only click on the links on the web page to browse the text, images, and other content set by the developer.
Soon, Web1.0 was replaced by Web2.0. Web2.0 is characterized by user-created personalized recommendations for interaction, where users are not limited to browsing the Web, but can also create their own content and upload it to the Web to share with others. The original purpose of Web2.0 was to bring the Internet closer to democracy and to make users more interactive.
Nowadays, the Internet has entered the Web3.0 era based on blockchain and artificial intelligence and marked by decentralization and intelligence \cite{hussein2014transition}. The Web3.0-based Internet has shifted banking business from offline to online, and digital transformation of the Internet has formed an industry consensual. And the development of web technologies could be illustrated in Fig. \ref{webevo}.

\begin{figure*}[h]
    \centering
    \caption{Applications of combining digital economy with Web3.0}
    \label{apps}
    \includegraphics[scale=0.22]{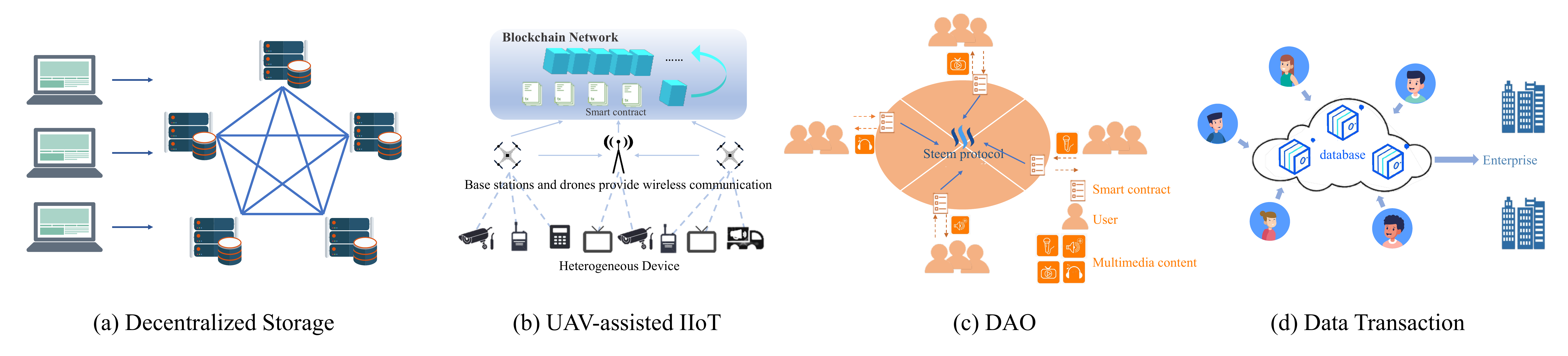}
\end{figure*}

In Web3.0, websites are allowed to have the ability to learn on autonomy themselves \cite{silva2008web}. Moreover, blockchain distributed storage technology is adopted to realize a decentralized autonomous network. Users can accomplish content publishing, economic transactions, and other actions without going through a centralized platform. 
They employ DAO to manage their digital identities, assets, and data by themselves, through the extended reality (XR) technology hardware and blockchain distributed storage technology together form the technical foundation of Web3.0 \cite{gadekallu2022blockchain}.
Therefore, Web3.0 can provide decentralized services for the digital economy, as well as unified consensus data valorization services, thus providing a favorable foundation for the development ecology of the digital economy. And some typical applications of combining digital economy with Web3.0 are illustrated in Fig. \ref{apps}.

In summary, the contributions of this paper are as follows:
\begin{enumerate}
	\item We introduce the preliminaries of Web3.0, and review the state-of-the-art studies of Web3.0.
	
	\item We present four core components of the digital economy, and then discuss how Web3.0 can be effectively integrated with the digital economy in each component.
	
	\item We envision typical challenges to shape the future digital economy in the next decades.
\end{enumerate}

The organization of this paper is as follows: Section II introduces the preliminary of blockchain, distributed storage, and privacy computing. Section III systematically introduces the related works on Web3.0 technologies. Some applications and challenges of combining digital economy with Web3.0 technology will be introduced in Section IV and Section V respectively.
Based on our discussion, the paper is summarized in Section VI.
\section{PRELIMINARY}
In this section, several fundamentals about Web3.0 will be introduced, such as blockchain, distributed storage and privacy computing.
\subsection{Blockchain}
\begin{figure}
    \centering
    \caption{Technology architecture of blockchain}
    \includegraphics[scale=0.28]{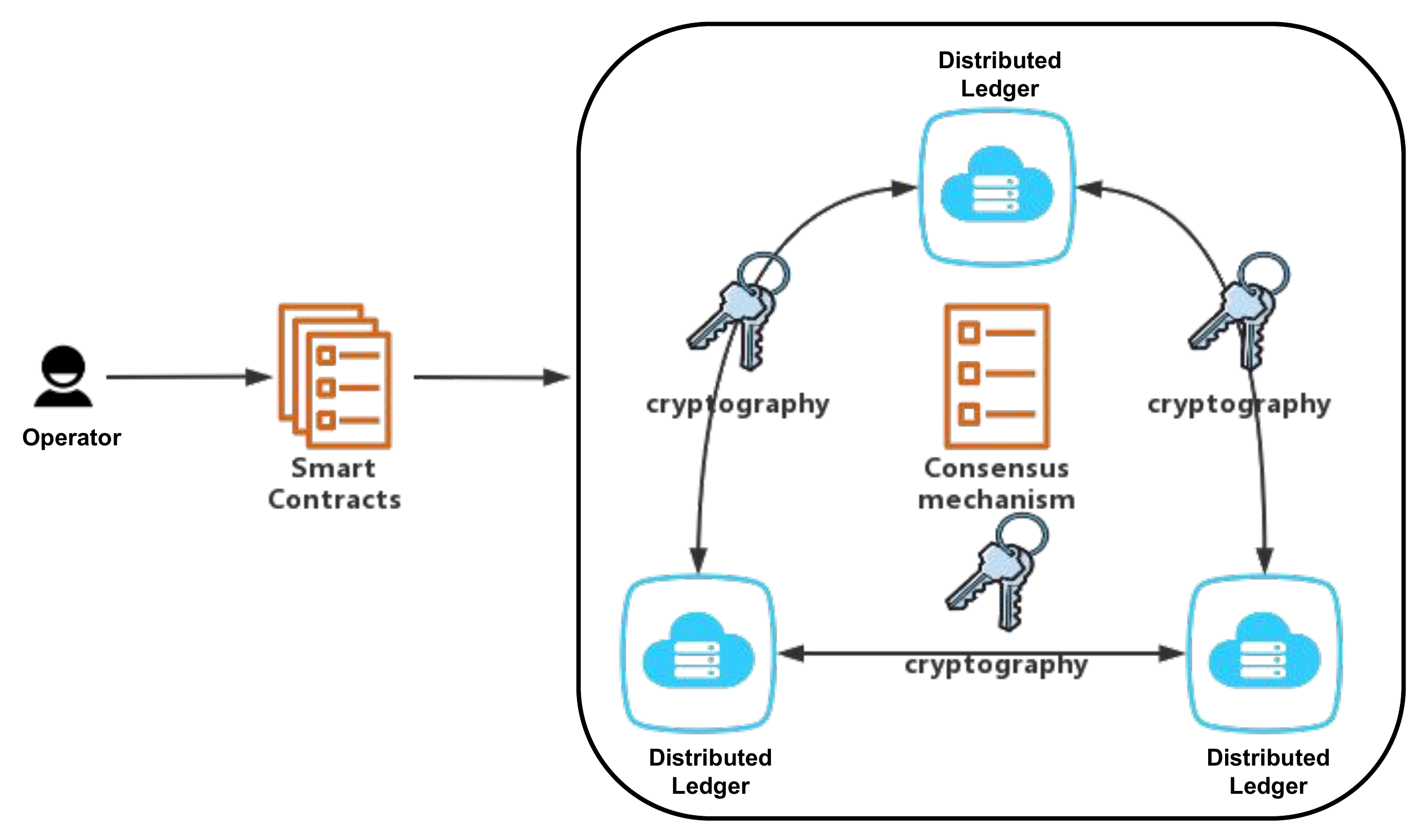}
    \label{blockchain}
\end{figure}

Blockchain is the main foundation of the decentralized system of Web3.0, and it returns digital sovereignty to users through the decentralized power of blockchain. 
The technology architecture of blockchain is given as Fig. \ref{blockchain}. The key technology stack\cite{zheng2019survey} will be introduced.
\subsubsection{Distributed Ledger}
The distributed ledger is a database on a peer-to-peer network without a central administrator that maintains data consistency among multiple enterprises or institutions through a consensus mechanism. Each node/enterprise/institution has a complete and identical copy of the data. Any changes or additions made to the ledger are reflected and copied to all participants in a matter of seconds or minutes\cite{rauchs2018distributed}. Distributed ledgers are inherently more difficult to be attacked because all of the distributed copies need to be changed simultaneously for an attack to be successful. As a result, cyber attacks are reduced using distributed ledgers.
And financial fraud is also reduced by the use of distributed ledgers because distributed ledgers provide for an easy flow of information\cite{hyvarinen2017blockchain}, which makes an audit trail easy to follow for accountants when they conduct reviews of financial statements. This helps remove the possibility of fraud occurring on the financial books of a company.
\subsubsection{Cryptography}
Since blockchain operates with a decentralized, peer-to-peer network model and nodes don't have to trust one another\cite{treiblmaier2021impact}. Therefore, blockchain must ensure appropriate safeguards for transaction information on unsecured channels while maintaining transaction integrity\cite{kosba2016hawk}. And consequently, cryptography becomes an essential requirement for blockchain to safeguard user transaction information and privacy alongside ensuring data consistency\cite{carrara2020consistency}.
Blockchain leverages the real-world concept of signatures by using cryptography techniques and encryption keys. Basically, cryptography is a technique for transmitting secure information between two or more participants\cite{katz2020introduction}. The whole process works by the sender encrypting the message with a specific type of key and algorithm and then sending it to the receiver. The receiver uses decryption to get the original message. Encryption keys ensure that unauthorized recipients or readers cannot read messages, data values, or transactions. They ensure that only the intended recipient can read and process a specific message, data value, or transaction. In recent years, many new tools related to cryptographic applications in blockchain have emerged with different functionalities.
\subsubsection{Consensus Mechanism}
In order to maintain the consistency of the distributed ledger, the blockchain needs to reach a consensus on the block transaction history. How to make each node keep its respective data consistent through a rule is a crucial problem. The solution to this problem is to develop a consensus mechanism.

The consensus mechanism is actually a rule according to which each node confirms its own data. In the blockchain systems, the most commonly used consensus algorithms include PoW, PoS, DPoS, and PBFT\cite{zhang2020overview}.
The blockchain system uses this consensus algorithm to make the ledger data of each node in the network agreeable.
\vspace{-2mm}
\subsubsection{Smart Contracts}
Smart contracts are decentralized, information-sharable program codes deployed on the blockchain that define the logic of applications on a decentralized network\cite{rouhani2019security}. Typically, they function as a digital protocol that follows specific rules for enforcement. These rules are predefined by computer code and are replicated and enforced by all network nodes.
Blockchain smart contracts support the creation of de-trusted protocols. This means that both parties to the contract make commitments through the blockchain without the need to know or trust each other\cite{lyu2019secure}. In addition, the use of smart contracts eliminates the need for intermediaries, thereby significantly reducing operational costs.
Smart contracts are autonomous and decentralized. Specifically, they usually automatically run the procedures defined in the smart contract when predefined conditions are met, without the intervention of any contract signatory. Besides, they not only give programmability to the underlying data of blockchain, but also encapsulate the complex behavior of each node in the blockchain network, providing a convenient interface for establishing upper-layer applications based on blockchain technology\cite{seijas2016scripting}, and therefore have a significant impact on blockchain.
However, when smart contracts are exposed to untrusted systems and lack government oversight, they still have security issues.

\subsection{Distributed Storage}
Distributed storage is the core technology that must be perfected first for the Web3.0 ecosystem and is an important cornerstone for building the underlying infrastructure of the Web3.0 ecosystem\cite{dimakis2011survey}. Distributed storage is based on blockchain technology that uses open source applications and algorithms to store sliced data in multiple independent network nodes. It advocates privacy protection, data backup redundancy, and value-oriented data by providing incentives for network nodes and content uploaders. Among the aforementioned features, the incentive model is an important aspect of distributed storage because it allows for long-term data preservation and security. Distributed storage technology in Web3.0 raises awareness of data security and user data ownership. The major distributed storage projects currently include BitTorrent, Filecoin, and Crust, where some of these projects will be introduced in the following subsections.
\subsubsection{BitTorrent}
BitTorrent is a decentralized transfer scheme proposed by Bramcoon in 2003. It uses an efficient software distribution system and peer-to-peer technology to share large files (such as a movie or TV show) and enables each user to provide upload services.
It did not require the content resource publisher to own the high-performance server to transfer the data, and the more users downloading the same file, the faster the download speed can be. 
In addition, the free model also attracts the use of the majority of Internet users.
However, BitTorrent must use the torrent file containing all targeted content addresses to perform the download. The downloading content is strictly restricted within the scope of the torrent file. 
Furthermore, BitTorrent lacked incentives to motivate users to share unpaid files.
So BitTorrent can be described as a prototype of the distributed storage model.
\subsubsection{Filecoin}
Filecoin is a content-addressable and peer-to-peer distributed protocol, defines how files are stored, retrieved, and transferred in a distributed system, and this enables permanent and decentralized preservation of files.
Besides, Filecoin is also a peer-to-peer network for storing files, with built-in economic incentives to motivate the behavior of various players in the network's storage and retrieval market, ensuring the safe and secure storage of files. Filecoin is built on top of IPFS to create a distributed storage marketplace for long-term storage.
\subsubsection{Crust}
Crust is an incentive layer protocol that implements distributed storage and is adapted to multiple storage protocols including IPFS. Besides, Crust is also known as the “Filecoin” on the Polkadot network, an incentive layer protocol based on the Polkadot parallel chain construct.
What makes Crust different from other distributed storage projects is its pioneering use of a hardware solution, Trusted Execution Environment (TEE) technology, as the core solution to quantify and verify the actual workload of nodes within the local CPU hardware.
Based on TEE, Crust proposed Meaningful Proof of Work (MPoW) to count the storage workload of nodes and report it to the chain.
Crust also proposes a PoS consensus algorithm that defines the number of storage resources, called Guaranteed Proof of Stake (GPoS). The workload report is recorded and packaged into a block along with other transactions to calculate a Staking amount, and then based on this amount, PoS consensus is performed.
\subsection{Privacy Computing}
Web3.0 emphasizes the protection of users' personal data, and therefore, as a key technology to solve the data privacy problem, privacy computing is becoming the immediate need of Web3.0 existence.
Privacy computing technology can analyze and calculate data under the premise of protecting data privacy and security, which provides a strong guarantee for efficient and safe circulation of data across industries and organizations.
Currently, privacy computing technologies are classified as Secure Multi-Party Computing, Federated Learning, and Trusted Execution Environment (TEE). In the following, we will introduce each of them.
\subsubsection{Secure Multi-Party Computation}
Secure Multi-Party Computation (MPC)\cite{evans2018pragmatic} was proposed by Andrew Chi-Chih Yao in 1982 through the Millionaire Problem.
It aims to solve the problem of collaborative layout for privacy protection among a group of participants who do not trust each other. Furthermore, it provides the data demanders with the ability of multi-party collaborative computing without disclosing the original data. MPC is mainly concerned with the problem of how to securely compute an agreed function without a trusted third party while requiring that each participant cannot get any input information from other participants except the computation result. It mainly involves zero-knowledge proof, homomorphic encryption, differential privacy, inadvertent transmission techniques, etc.
However, the higher computational or communication complexity puts some limitations on the usability of MPC.
\subsubsection{Federated Learning}
\begin{figure}
    \centering
    \caption{The workflow of federated learning}
    \label{workflowoffl}
    \includegraphics[scale=0.45]{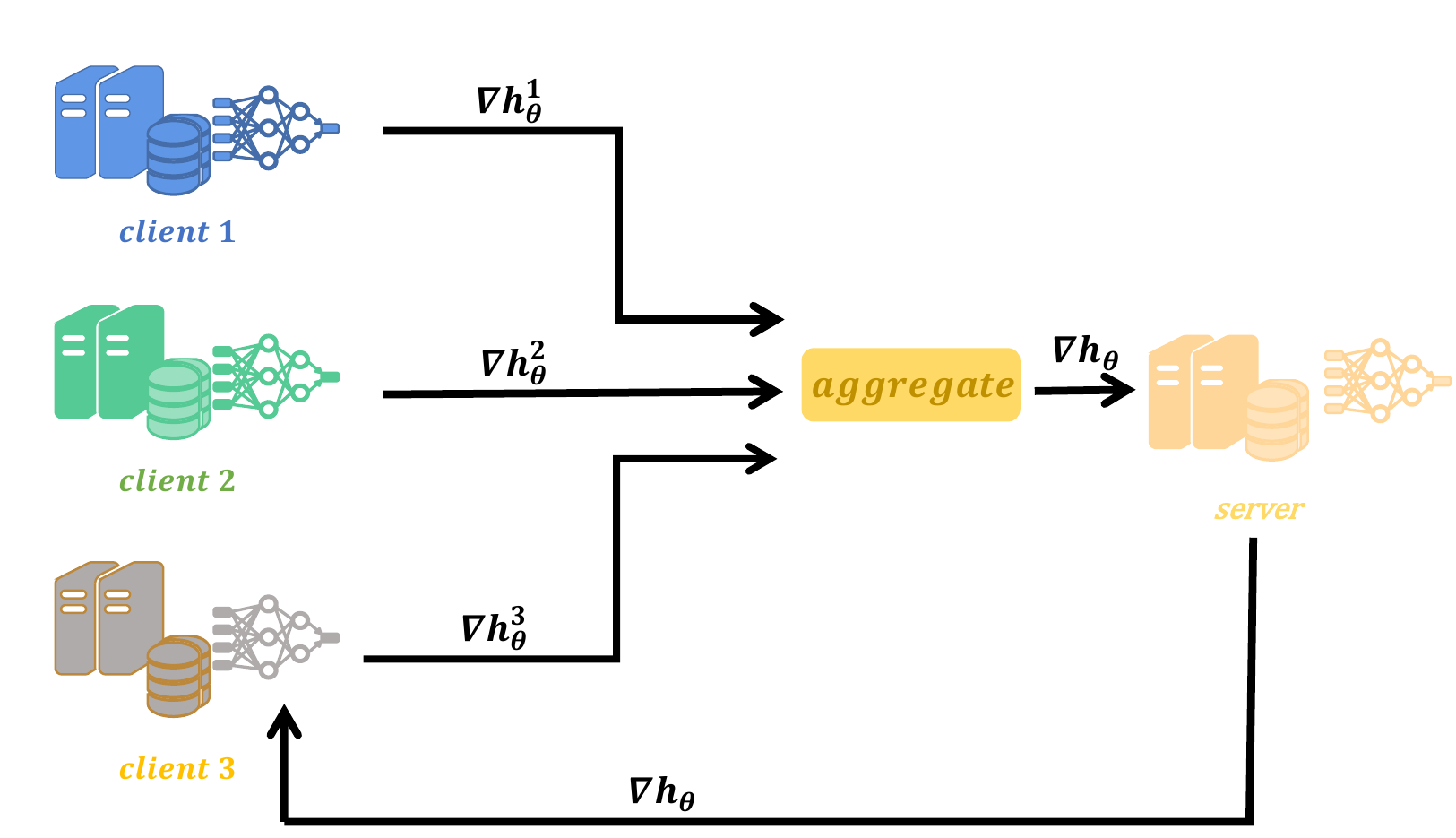}
\end{figure}

Federated learning is a machine learning technique developed to solve the problem of data silos\cite{li2021survey}. Its goal is to carry out efficient machine learning among multiple participants or multiple computing nodes while guaranteeing security and protecting privacy when exchanging data. Currently, federated learning is classified into three categories according to the different data distributions among participants: horizontal federated learning, vertical federated learning, and federated transfer learning.
The workflow of federated learning is given in Fig. \ref{workflowoffl}.
\subsubsection{Trusted Execution Environment}
The Trusted Execution Environment is based on a hardware-secure CPU that implements memory-isolated secure computing, allowing privacy-preserving computations to be performed with guaranteed computational efficiency\cite{sabt2015trusted}. 
The basic idea is that all calculations of sensitive data are performed in an isolated memory that no other hardware parts can access except for authorized interfaces. In this way, private computation of sensitive data is achieved.
The security comes from the isolated hardware device's ability to resist attacks while avoiding additional communication processes and significant computational overhead. However, the disadvantage is that its security relies heavily on the hardware implementation, making it
vulnerable to attacks from different attack surfaces.
\section{RELATED WORK}
Recently, Web3.0 has been widely researched, and more representative technologies will be introduced.
\subsection{Fundamental Research}
The crucial component of Web3.0 is the network structure and security is important nature of Web3.0. In order to enhance the security of the network to meet the requirements of Web3.0, there are several efforts to give their respective solutions.

Nowadays, web applications are built on centralized network protocols, and these protocols have disadvantages such as being vulnerable to attacks and prone to single points of failure. And therefore, Mainframe\cite{clarke2018mainframe}, a fully decentralized communications layer, is proposed to replace the original centralized network protocols. Based on this, many applications could be easily extended to decentralized scenarios, such as email service.

According to the definition of Web3.0, blockchain is a core component and could be regarded as a backbone where Web3.0 applications are built, and where the data is stored. For privacy issues, the data should be encrypted and stored in a distributed manner with multiple copies. However, traditional solution proof-of-replication (PoRep) consumes a lot of resources. For reducing the consumption of resources, Kan \textit{et al.} proposed a new PRE scheme\cite{kan2022proxy}, in which transferring the ciphertext into a new one is not required, and therefore conserving computing resources.

Recently, blockchain technology has also been widely researched, and Huang \textit{et al.}\cite{huang2021survey} reviewed the research related to blockchain, and then gave several major research directions. In terms of smart contracts, Zheng \textit{et al.}\cite{zheng2020overview} reviewed the research related to smart contracts, and Kong \textit{et al.}\cite{kong2022characterizing} proposed a novel method to characterize and detect gas-inefficient patterns. Considering smart contracts are vulnerable to attack, SmartDagger\cite{liao2022smartdagger} is proposed to detect cross-contract vulnerability based on the bytecode of smart contract. To make blockchain technology healthier, Park\cite{zheng2022park} used parallel-fork symbolic execution to accelerate smart contract vulnerability detection. Considering the lack of regulation of blockchain, the Ponzi schemes on Ethereum\cite{chen2018detecting} should be detected, and permissions\cite{zheng2022aeolus} should be appropriately limited.
\subsection{Benefits to Artifical Intelligence}
Traditional artificial intelligence technologies are typically presented in a centralized form, however, this generally leads to inflexibility, poor scalability, etc.

Cao \textit{et al.} presented a new concept named Decentralized Artificial\cite{cao2022decentralized}, and the authors also give the main research areas where Web3.0 and artificial intelligence are combined, such as Edge Intelligence and Decentralized Communication.

With Web3.0, artificial intelligence could present a decentralized form, and data could be stored by different parties. However, relevant laws and regulations, such as GDPR, may constrain the transmission of data, which has a significant impact on AI technologies that require large amounts of data to support them. To make data available in privacy-protection scenarios and enjoy the benefits of decentralization, federated learning\cite{yang2019federated} is proposed. In the federated settings, models with local data characteristics replace the original data for transmission and aggregation, and after several rounds of training, a global model with all data features is obtained. For mobile edge computing, combining blockchain with deep reinforcement learning, the performance of computation offloading could be increased\cite{qiu2019online}, and the bitcoin network could be used to enhance complex network analysis\cite{tao2021complex}.
\subsection{Benefits to Education}
COVID-19 makes it impossible for many educational institutions to conduct face-to-face sessions, leading online education especially important. Jiang\cite{jiang2014will} believes Web3.0 could bring a new environment for online education and provide personalized technologies to enhance learning effectiveness. 

To improve learning outcomes, an online 3D meeting application\cite{gupta20223d} based on Virtual Reality (VR) and Web3.0 is developed. A holographic-based approach is used to ensure intuitive learning, and to bridge the remote monitoring gap.

For academic staff or students in higher education, subject information and resources are vital in teaching activities and scientific research. However, these resources are stored separately, and this also poses a challenge for information integration. Considering Web3.0 is also known as the semantic web where knowledge is connected, bringing the ideas of Web3.0 for information integration could be a good idea for information integration. Therefore, Han \textit{et al.}\cite{xiaoting2010subject} proposed to build a subject information integration system using the new technologies of Web3.0.
\subsection{Benefits to Data Management}
Recently, the generation of large amounts of data has increasingly tended to be decentralized, and this poses a great challenge to data management. Fortunately, Web3.0 technologies could be used for the management of large amounts of decentralized data.

IoT devices should be controlled by a third party, and they usually require for transmitting sensitive user data\cite{fernandes2016security}. Therefore, Ayoade \textit{et al.}\cite{ayoade2018decentralized} proposed a decentralized system of data management for IoT devices, where all data access privileges are stored in the blockchain, and smart contracts are used to manage interactions between devices. To make the system safer, TEE is used to store the row data.

When using blockchain to construct data management systems, storage should receive more attention. Since the former data could not be deleted, the storage costs can increase significantly. InterPlanetary File System could be used to solve this problem\cite{poornima2021secure}. When storing data into blockchain, the hash value is uploaded, rather than the row data, and the hash value could be acquired by uploading data into IPFS. Considering the hash value is smaller than data itself, storing the hash value decreases costs. And to enhance the security of data storage, MOOCsChain\cite{li2022moocschain} is proposed to incorporate blockchain to enhance security.
\subsection{Benefits to Business}
Web3.0 could be applied to the financial domain, and may change the way how companies use the collected information and sell their products. Almeida \textit{et al.}\cite{almeida2014commerce} analyze how Web3.0 affect business and give nine kinds of potential business models.

Business models are explored from many aspects. Momtaz \textit{et al.}\cite{momtaz2022some} believe Web3.0 may give rise to new products and business models since the components of Web3.0 reduce transaction costs and the trust of interaction between social and economic has been formed for the decentralized consensus mechanisms.

Toyoda \textit{et al.}\cite{toyoda2022web3} apply Web3.0 to behavioral economics, and propose an incentive mechanism based on crypto-enabled services, and this mechanism is general, and even can be applied for services that required making decisions under an uncertain environment. In order to prevent from internet giants monopolizing the power to use user data, Web3.0 needs to establish a decentralized identifier (DID), and the DID could be used to link user data in the form of DID documents. Hence, user assets also need to be represented decentrally. Non-Fungible Token (NFT)\cite{wang2021non} is proposed to represent physical assets in a decentralized form, which have become an important part of Web3.0. However, the development standards of NFT are missing, leading to a couple of underlying systems and failing to govern physical asset value mapping. Yang \textit{et al.} proposed a general NFT architecture for Web3.0\cite{yang2022generic}, and they used a universal connector to connect the upper application environment and the underlying value mapping of the physical asset environment.

Based on Web3.0, an online shopping platform\cite{xiao2017design} is designed. The platform incorporates mainstream technology into artificial intelligence and visualizes the 3D presentation of products to ultimately enhance the shopping experience. These mainstream technologies include Web3D, Augmented Reality (AR), and so on.
\section{DIGITAL ECONOMY UNDER WEB3.0}
\par Web3.0 is considered as an emerging way of how to organize an internet structure. Many internet and industrial applications are now obeying the paradigm of server-client structure. Unlike most of the above applications, the new decentralized Web3.0 applications like cryptocurrency provide many systems with a new organizing method to satisfy the new meets of security, parallel, and other requirements in recent internet environments.
\par In this section, firstly we will introduce the conception and the classification of digital economy and then describe all these categories by analyzing the existing technologies and applications, also by giving an intuitive case as an example in each subsection. 
\par The detailed classification of digital economy and its conception could be divided into four parts as follows:
\begin{itemize}
    \item Digital Industrialization: The industry includes all the products, hardware, and software around the Information technology industry. This is the most important topic where most recent applications using Web3.0 appeared. In this part, some new technologies like cryptocurrency, decentralization social software, and metaverse software will be introduced.
    \item Industry Digitization: Industry digitization is committed to using advanced information and digital technology to accelerate innovation in traditional industries like agriculture to help achieve better efficiency and progress during the production process. The main point in this topic often focuses on designing an industrial internet to implement better cooperation during the production process.
    \item Digital Governance: Digital Governance helps all kinds of cooperation agencies to reestablish their rule-based management system by combing with digital technology and also provides assistance to public service.  Due to the traditional rule or contract having borne the latent risk of unforeseen and unordered humankind activity like rule-breaking, the introduction of digital technology aims to reduce the influence of these kinds of manners and creates a cooperation system mainly by agreements formulated previously and executed strictly by the machine. 
    \item Data Valorization: Most research on data science attempt to augment the value of data, especially in ways like machine learning. Federated learning is a typical way of using a distributed way to carry out the process of machine learning and satisfy the demand for data security and other need like calculation efficiency and privacy or so. The main topic of data valorization is finding out a method to collect and fully utilize the everyday data produced by each individual.
\end{itemize}
\par
\begin{figure}
    \centering
    \caption{The relation between digital economy and Web3.0}
    \label{relation}
    \includegraphics[scale=0.24]{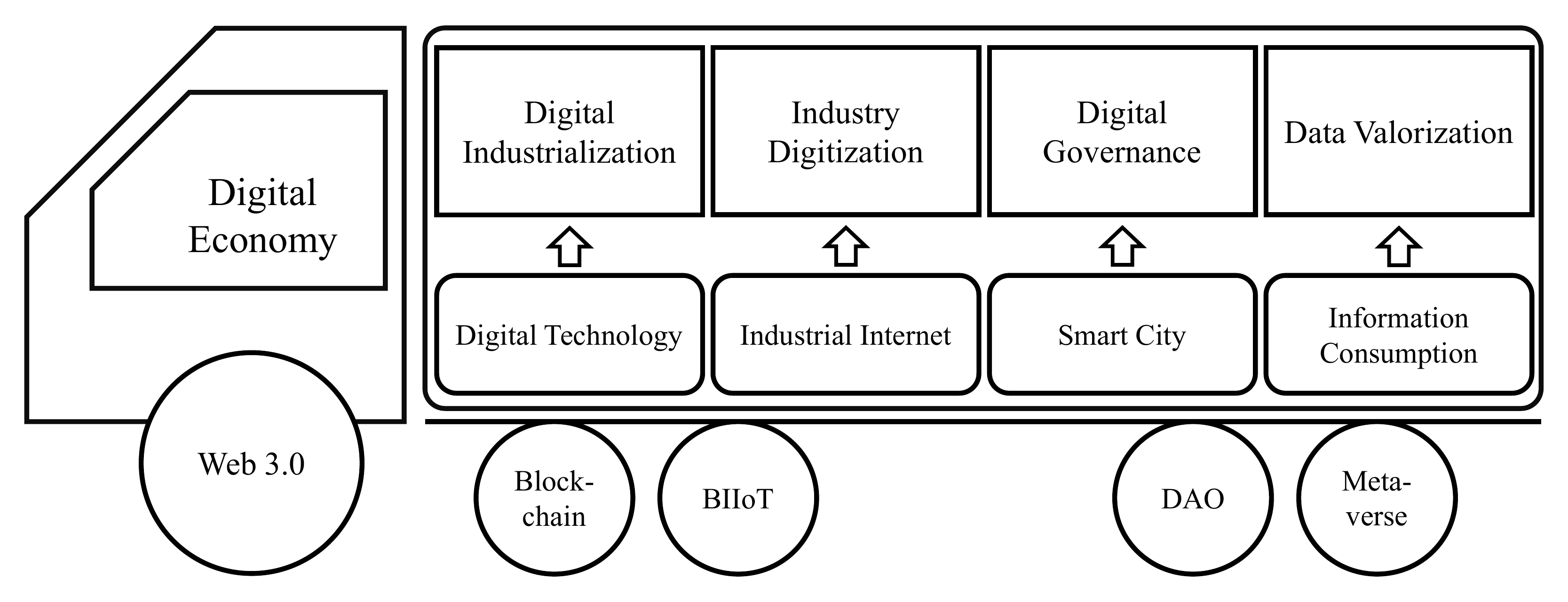}
\end{figure}

 The main Web3.0 applications and their connections with digital economy are listed in Table \ref{applications}. And the relation between digital economy and Web3.0 could be illustrated in Fig. \ref{relation}. As this figure illustrates, Web3.0 could promote the development of the digital economy as an engine. Taking digital industrialization as an example, with the help of blockchain, digital technology could be presented as a decentralized form, and therefore enhance the security of the digital economy environment.
\begin{table*}
    \centering
    \caption{Web3.0 with Digital Economy}
    \label{applications}
    \begin{tabular}{|c|c|c|c|}\hline
    & Blockchain              & Decentralization            & Metaverse          \\ \hline
    Digital Industrialization & DeFi,Cryptocurrency,NFT & Decentralized Storage Technology & Metaverse Software \\ \hline
    Industry Digitzation      & IIoT                   & M2M mechanism                 & AR       \\ \hline
    Digital Governnace        & Smart City              & DAO                       & -              \\ \hline
    Data Valorization         & Federated Learning      & Data Storage                   & Metaverse with AI \\ \hline
    \end{tabular}
\end{table*}
\subsection{Digital Industrialization}
\par Web3.0 is an emerging way of network architecture organization. The applications based on it naturally impact the related industries of Internet and information technology. With the adoption of a novel design of the blockchain system, Nakamoto first designed a digital currency Bitcoin\cite{nakamoto2008bitcoin}, which opened the prelude to the transformation of the digital economic system, also described as Decentralized finance. The key feature of blockchain has brought this cryptocurrency a lot of benefits on persistency, anonymity, and auditability\cite{weyl2022decentralized}. After that, various cryptocurrencies like Litecoin\cite{haferkorn2014seasonality}, Ethereum\cite{chen2020survey}, and USDT\cite{frankenfield2019tether} have emerged one after another and put forward a new idea for the new situation of currency under Web3.0 to satisfy the new need for a digital currency. In addition to this, blockchain has become a solution for many other applications due to its features of decentralization, privacy, and security preserving. The Smart Red Belly Blockchain system\cite{tennakoon2022smart} created an enhanced transaction management system for the decentralized applications to use. In this newly designed blockchain system, the author provides all the decentralized apps with faster transaction execution. Also, the hyperledger fabric project\cite{cachin2016architecture} gives the developer a powerful tool to develop new applications based on blockchain. This project has given a very clear and direct definition of blockchain which can be seen as a cooperative ledger organized in a decentralized way. Based on this SDK, developers can use whatever programming language to implement a blockchain system on their own. 
\par Another important application of blockchain in Web3.0 is a Non-Fungible Token (NFT). NFT almost like its name creates a non-fungible mark on the chain to enable artists to create works that can be traded with their digital certificates. According to the definition of NFTs, NFTs represent the rights to unique digital assets and are presented in digital form, and NFTs could even be traded on the market using blockchain technology\cite{wilson2021prospecting}.

The metaverse is an aggregated vessel where lots of technologies can be used to build a virtual world. Therefore, a lot of research put their eye on how to apply Web3.0 technology into the metaverse\cite{yang2022fusing}. 

\par Storage is another important content in the network infrastructure. Given that most of the current storage is collective, the Web3.0 tech gives a new consumption of constructing a decentralized storage system. We may be familiar with this form of storage like the P2P file sharing system BitTorrent protocol\cite{erman2022bittorrent} early in 2003. This storage form without a central server may bring out many challenges, but it's a vital question for real decentralized applications, considering the data in these apps must be decentralized too. The technology route of decentralized storage is divided into two main parts according to whether it is based on blockchain or not. Like the IPFS protocol, we discussed the previous chapter and the application Filecoin\cite{bauer2022filecoin} based on it. The article\cite{trautwein2022design} analyzes the IPFS storage protocol function as a layer of Web3.0 storage. Also, there exist some storage systems relying on blockchain like Arweave\cite{williams2019arweave} which is similar to Ethereum to store the data on the chain and use the token as a reward to motivate the miner storing files on the chain. Many decentralized applications now still rely on a centralized client for simplicity and usability, so storing the data decentrally will be a critical point to the coming of the real decentralized Web3.0. 
\subsection{Industry Digitization}
The main infrastructure of industry digitization is industrial internet, and the design of industrial internet aims to connect machines to networks and harvest the production data while also being able to control them remotely, which finally achieves the goal of combing software and machines together. 
The abstraction of software connects the physical world and can optimize the whole factory above the scale of a single machine. Some entity assets will be replaced by online software and human labor can also be liberated by the productive forces of a highly connected internet of machines.
\par To build an industrial internet, it should be considered how to connect machines with the upper layer of internet. Internet of things\cite{rose2020internet} has provided a good research base for the communication between physical objects with the networks. The emergence of blockchain naturally gives these distributed machines a good way to organize an industrial internet of things (IIoT) as the basis of industrial internet\cite{ahmad2019blockchain, chen2019cooperative}. To implement a reliable IIoT, some key features like interoperability, heterogeneity, confidentiality, and safety vulnerabilities must be solved with a reliable plan. And some of these challenges that arise in IIoT can be overcome by using a current form of blockchain techniques that provides solutions with enhanced safety, and dependability\cite{kumar2022survey}.
\par Some component has been used in the industrial internet already and has shown a great influence in the blockchain-based IIoT. A novel communication model, named M2M \cite{wu2017m2m,meng2016data}, uses the blockchain to connect the machine with the machine and integrate the factorized machine. Based on this system, some design a power management system, in which the power is transacted between the generator and the machine and can also automatically adjust the load state of the different grids of the machine. The unmanned aerial vehicles (UAV) assisted M2M system\cite{li2020uav} further enhanced the stability of the IIoT, which makes a more capable system of data computation and decision-making.
\par In addition to these specific models, the Web3.0-formed IIoT has already been used in various industries. For example, the blockchain has already been used in the food supply chain as an effective way of monitoring the quality and safety of food by providing a more traceable and transparent chain of food production\cite{8290114,pena2019blockchain}. The whole supply chain contains from production to transportation and meanwhile involves many subjects like enterprise, consumer, and logistics, which gives the traditional way of monitoring lots of issues that is hard to solve\cite{lin2019food}. Smart Farming is another important field in which the factors impacting agriculture are complicated and messy and bring many challenges. Fishing now uses many sensors which can provide many data like temperature, humidity, water level, etc. Blockchain can help integrate these decentralized data and helps farmers to monitor and control\cite{ hang2020secure}. Other industries like manufacturing\cite{ abeyratne2016blockchain , leng2020blockchain }, petrochemical industry\cite{lu2019blockchain, ahmad2022blockchain }, automotive manufacturing\cite{ guhathakurta2018blockchain , sharma2018blockchain } also use the blockchain as a plan to collect and analyze the production data. 
\par Another possible form of the industrial internet is complemented by metaverse. Metaverse aims to communicate the 3D-formed virtual worlds with the real physical worlds\cite{li2022internet}. With the help of AR and VR, many applications can be implemented in the world of metaverse like the healthcare\cite{tan2022metaverse}, the social software\cite{ jiaxin2022socializing}, the entertainment\cite{ kye2021educational }, and the smart city construction\cite{ yoo2022digital}.

\subsection{Digital Governance}

Decentralized Autonomous Organization (DAO) is an important method to achieve digital governance, enabling people to coordinate and govern themselves mediated by a set of self-executing rules deployed on blockchain. The governance form of DAO is decentralized\cite{hassan2021decentralized}. The early-stage application of DAO like crowdfunding is a specific example with the problem that small-scale investors may suffer from financial mismanagement and outright fraud. The blockchain system Ethereum\cite{ wood2014ethereum} may have the possibility to solve the issues above. Ethereum acts as an intermediary between the participants and the token, which represents the profit and the right of the participant in this DAO.

Recent community is committed to exploring the other usage of the blockchain and its token of DAO. The stem is a social media platform, it uses the “stem” token to reward the user who uploads the content of text, image, video, or live stream which receive “likes” from others. In this form of social media, the benefits are owned by the content generator instead of the platform, making the participants both users and owners\cite{ chang2019understanding}. Another application is Augur\cite{peterson2015augur} based on Ethereum to create a market forecasting system.
\subsection{Data Valorization}
Data Valorization analyzes the daily produced data and makes them benefit economically. And the progress of data valorization is often divided into four steps: data requirement and labeling, data analysis, data storage, and data transaction. Due to the limitation of the calculation capability and the storage, putting all these progress distributable often becomes a good way. Also decentralizing all these progresses helps the acquisition of data from the number of users, the calculation using enough computing power, and the storage of data.

Federated learning\cite{zhang2021survey, huang2022contextfl} studies the distribution machine learning program which includes data communication, privacy security, and making good use of the amount of data to train a reliable model. Obviously, the nature of the blockchain can provide the federated learning system with some key components, and satisfy some vital requirements of federated learning\cite{li2022blockchain}. Fed2Coin\cite{liu2020fedcoin} model implements the federated learning system by using blockchain to help overcome the difficulty of the calculation of Shapley Value, the contribution of each user in the cooperation. Fed2Coin helps solve the profit distribution problem among all the participants during the data valorization. Besides, the model BAFFLE\cite{ramanan2020baffle} aims to design an aggregator based on blockchain. Considering federated learning systems are vulnerable to attacks, BFLC\cite{li2020blockchain} is proposed to defend against byzantine attacks, and proved to be convergent\cite{che2022decentralized}.
\section{CHALLENGES}
\subsection{Web3.0 for Digital Industrialization}
Digital industrialization is primarily to promote digital technology to form a large-scale industry.
Currently, Web3.0 is based on the concept of decentralization and the application of blockchain and other digital technologies to create a new digital ecosystem that integrates multiple scenarios into one. Besides, Web3.0 is an Internet infrastructure owned and trusted by users and builders. Therefore, in the role of Web3.0, digital technology based on blockchain and other digital technologies promote information technology services and consequently accelerates digital industrialization based on digital technologies.

In the era of big data, organizations tend to collect as much data as possible, which is prone to the problem of user privacy violation and threatens the data security of enterprises and individuals\cite{jain2016big}.
Undeniably, blockchain technology has many advantages in terms of privacy protection. However, the existing technology is still in the stage of development and improvement, while the characteristics of blockchain itself can no longer meet the user's demand for privacy protection.
Anonymity is a critical feature of blockchain technology, but this feature brings conflict with privacy protection\cite{gille2021limits}. 
In addition to privacy issues, Web3.0 is a technology that is considered impractical and expensive.
However, blockchain-based Web3.0 systems are extremely cost-inefficient compared to centralized systems such as Amazon Web Services, which can only process a few transactions per minute\cite{guo2016blockchain}.
Since Web3.0 can truly achieve the “trustworthiness” of blockchain, it must achieve consensus across the network, which will inevitably affect the transaction throughput.
Due to the inefficient PoW algorithm of the blockchain, it consumes a large amount of energy when we save data on the blockchain\cite{xu2018making}. Thus Web3.0 tends to be energy intensive, technically wasteful, and to handle only a limited amount of data.
\subsection{Web3.0 for Industry Digitization}
Industry digitization is mainly to use digital technology to support and promote the transformation and upgrading of traditional industries, while Web3.0 provides a new scene for this process and can accelerate the growth of information consumption, thus promoting industrial digitization. During the process of industry digitization, data is a key component. And therefore, efficient data management technology in Web3.0 should be paid more attention to make data work better.

In the context of Web3.0, the design of the industrial internet is now facing the issues like communication among heterogeneous devices under the circumstance of calculation resource constraints. The example of the previous UAV-assisted IIoT is a way to fix this by using a different strategy of deployment of the base station, especially when the devices of the industry are often located fragmentedly where the communication condition is poor. Research on edge computing\cite{cao2020overview} may help resolve the heterogeneity. Another imagination of the industrial internet is the digital twin\cite{tao2018digital} using the metaverse technology which is now still under exploitation. The problem of practicalizing the metaverse lies in the cyber world's scalability, and the computer graphic's bottleneck\cite{li2022internet}.
\subsection{Web3.0 for Digital Governance}
Lawrence Lessig argues in \textit{CODE}\cite{lessig2000code} that the network order will be regulated by laws, codes, markets, and codes at the same time.
With the network operator as the main target of regulation, the thought behind it is to require the operator to have the ability to be responsible for the data in the server, which is the traditional regulatory thought in the era of network centrality. Currently, digital governance is regulating blockchain information service providers as well as the filing system, without really considering the challenges that decentralization brings to regulation, let alone the challenges that DAO operations bring to regulation.

Decentralization is the essential feature of blockchain technology, and the existing traditional regulatory model is the exact opposite of this feature,  which will inevitably bring great difficulties to digital governance.
First, for the purpose of promoting the development of technology, timely and effective regulation of the technology itself is difficult. Besides, the cost of regulation is excessive. Blockchain technology was born among a group of anarchists called “cyberpunks”, and the “decentralization” has led to the fact that the privacy on the chain is no longer centrally and uniformly managed, but held by the users of each node. In this process, the scope of regulation is not clear, and the subject of regulation cannot be identified, so it is difficult to counter the “dishonesty” of the technology itself under the cloak of the “trustworthiness” of Web3.0.
For example, one of the earliest and most successful applications of blockchain is Bitcoin, which was in a sense created with evil intentions. Admittedly, Bitcoin is widely used in the “dark web” as a way to launder money and illicit transactions\cite{li2019toward}, as well as a tool to fund terrorists and insurgents. Therefore, while maintaining the advantages of the blockchain, integration into real-world regulatory systems is a necessary path to the widespread adoption of Web3.0.
\subsection{Web3.0 for Data Valorization}
Considering the decentralized feature of Web3.0, the five core progresses of data valorization, data collection and labeling, data analysis, data storage, and data trading, have the potential for decentralization, thus breaking through the limitations of computation and storage. However, in the process of decentralization, many problems may arise.

When building a blockchain-based federated learning system for data analysis, the client's model is uploaded to the blockchain's distributed ledger in the form of a transaction. However, considering the storage as well as consensus efficiency issues, a limit is generally imposed on the size of a single transaction, while in practice, the client-side model may adopt a very large model, resulting in exceeding the blockchain's constraint on transaction size\cite{shah2021model}. Although reducing the size of the model as much as possible can alleviate this problem to a certain extent\cite{cui2020creat}, it not only cannot solve the problem fundamentally but also may have the side effect of decreasing the effectiveness of the model and the accuracy of data analysis. Therefore, more efficient consensus algorithms need to be researched. As the model training process proceeds, more and more models will be saved to the blockchain. When model aggregation is performed on the blockchain, the query time will increase when there are more models on the blockchain, because all client models uploaded in this round need to be queried.
\section{CONCLUSION}
Web3.0 technologies are expected to play an important role in the digital economy. For example, blockchain technology could be used to enable decentralized data storage while combining with federated learning to solve possible privacy problems in the process of data analysis and can guarantee the transparency and fairness of data trading environment. Combined with the relevant technologies of Web3.0, the pain point problems in the digital economy will be solved, and the latest applications can be developed rapidly.

By investigating the most relevant work of Web3.0 in artificial intelligence, education, data management, finance, and Web3.0-based technologies, we summarize technologies that can be applied to the four core components of the digital economy, and provide Web3.0-based solutions to the problems existing in the development of the digital economy. Besides, we also analyze the critical challenges and unresolved issues that may arise in deeply integrating Web3.0 with the digital economy, and point out the direction for future research and application in this area for a good development ecology of the digital economy.

\bibliographystyle{abbrv}
\bibliography{web3.bib}

\begin{IEEEbiography}[{\includegraphics [width=1in,height=1.25in] {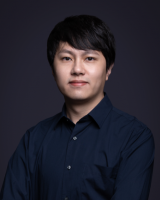}}] {Chuan Chen} \, (Member, IEEE) received the B.S. degree from Sun Yat-sen University, Guangzhou, China, in 2012, and the Ph.D. degree from Hong Kong Baptist University, Hong Kong, in 2016.  He is currently a Research Associate Professor with the School of Data and Computer Science, SunYat-Sen University. His current research interests include graph neural network, trustworthy machine learning, and financial big data. He published over 70 international journal and conference papers, including 1 ESI highlycited papers. He currently serves as the Associate Editor of the journal Software Impacts.
\end{IEEEbiography}

\begin{IEEEbiography}[{\includegraphics [width=1in,height=1.25in] {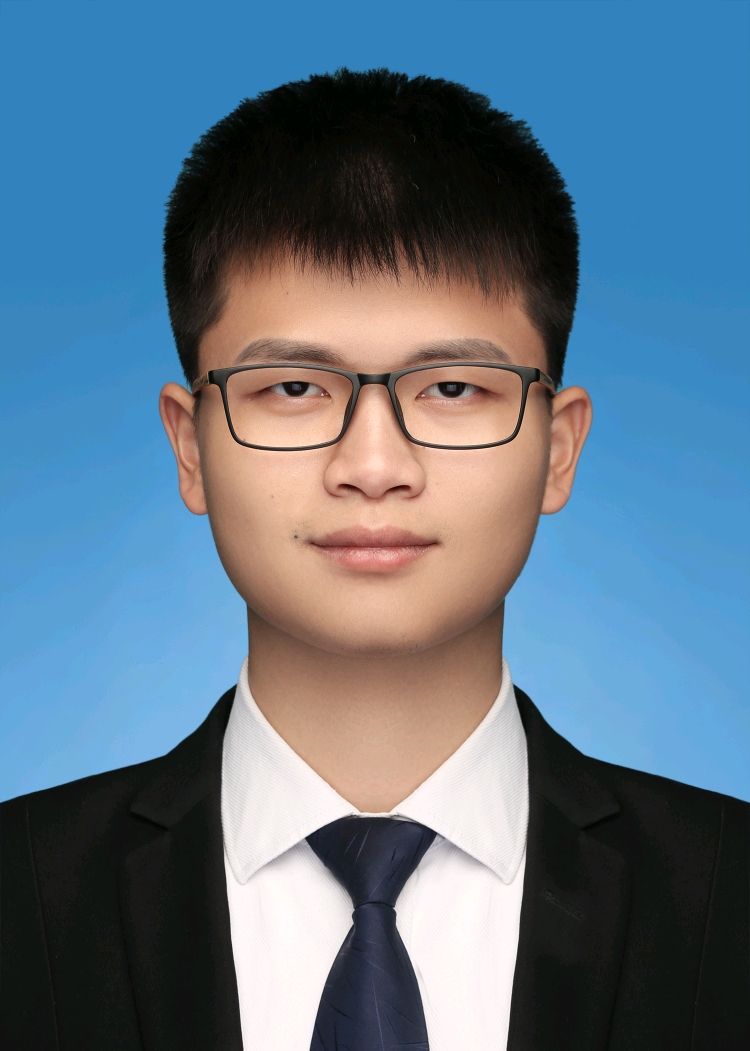}}] {Lei Zhang} \, (Student Member, IEEE) received the B.S. degree in computer science from Anhui University, Hefei, China, in 2018, and continues studying as a graduate student in computer technology in Sun Yat-sen University. His research now focus on federated learning.
\end{IEEEbiography}
\begin{IEEEbiography}[{\includegraphics [width=1.25in,height=1.25in] {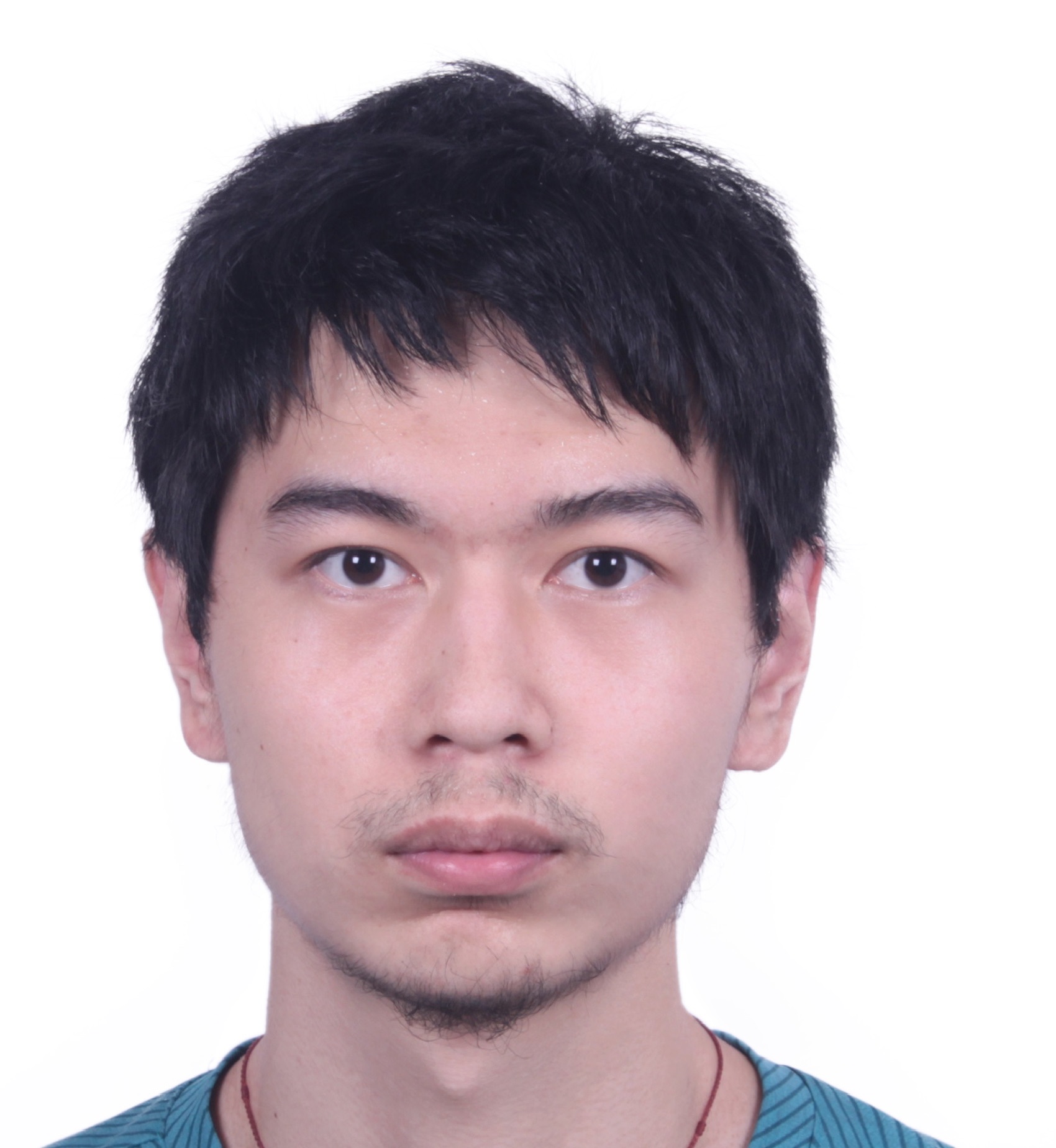}}] {Yihao Li} \, (Student Member, IEEE) received the B.S. degree in information engineering from Sun Yat-sen University, Guangzhou, China, in 2022 and continues studying as a graduate student in computer technology in the same university. His research now focus on the federated learning.
\end{IEEEbiography}
\begin{IEEEbiography}[{\includegraphics [width=1in,height=1.25in] {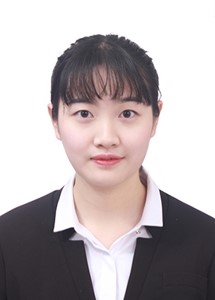}}] {Tianchi Liao} \, (Student Member, IEEE) is currently working toward the master's degree with the School of Computer Science and Engineering, Sun Yat-Sen University, Guangzhou, China. Her research interests include machine learning and federated learning. 
\end{IEEEbiography}
\begin{IEEEbiography}[{\includegraphics [width=1in,height=1.1in] {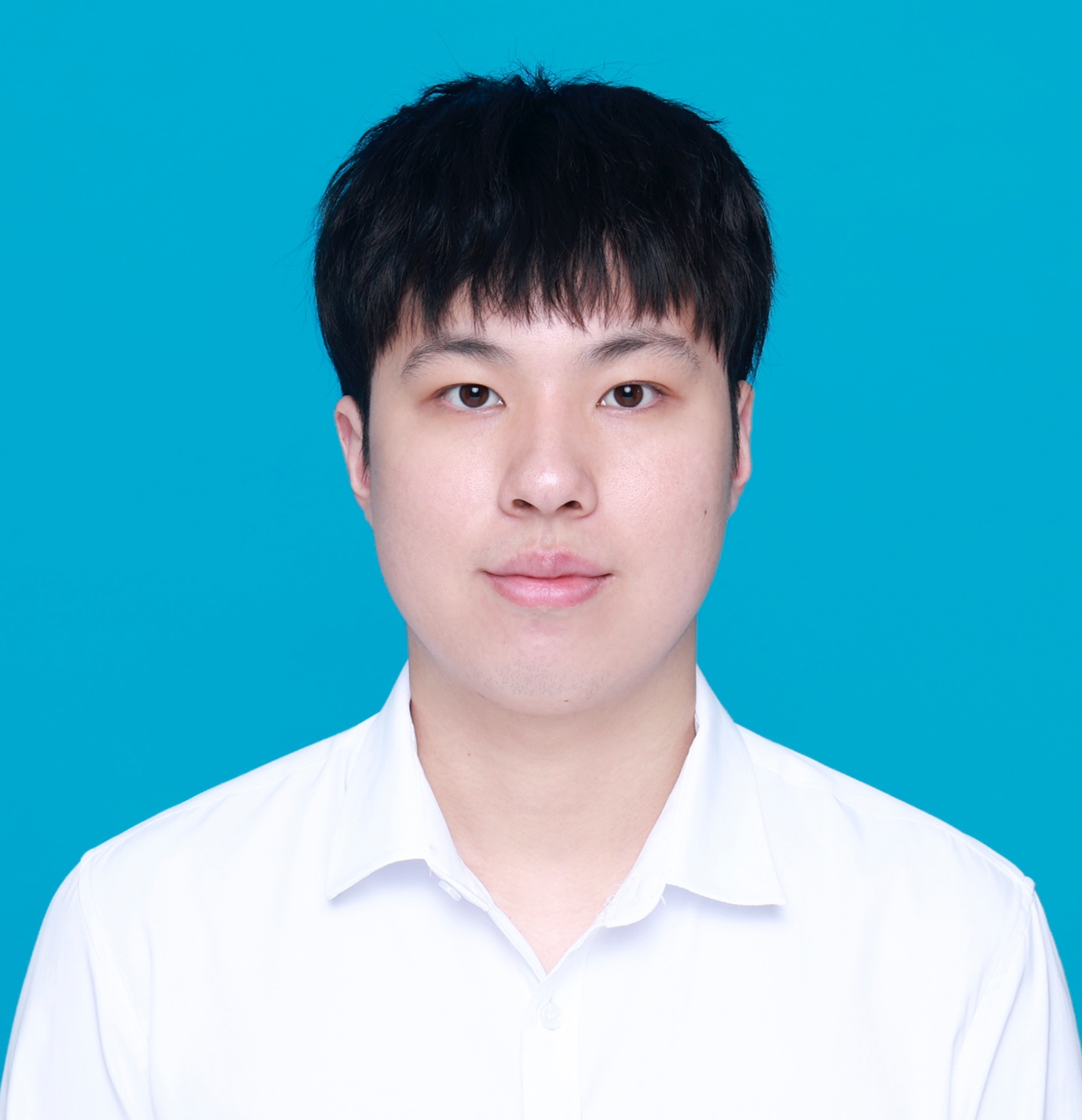}}] {Siran Zhao} \, (Student Member, IEEE) received the B.E. degree in software engineering from Central South University, Changsha, China and is currently working toward the master's degree with the School of Computer Science and Engineering, Sun Yat-sen University, Guangzhou, China.  His research interests include federated learning and graph machine learing.
\end{IEEEbiography}
\begin{IEEEbiography}[{\includegraphics [width=1in,height=1.25in] {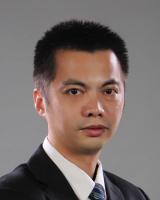}}]{Zibin Zheng} \, (Senior Member, IEEE) received the Ph.D. degree from the Chinese University of Hong Kong, in 2011.
    
    He is currently a Professor at School of Data and Computer Science with Sun Yat-sen University, China. He serves as Chairman of the Software Engineering Department. He published over 120 international journal and conference papers, including 3 ESI highlycited papers. According to Google Scholar, his papers have more than 7000 citations, with an H-index of 42. His research interests include blockchain, services computing, software engineering, and financial big data. He was a recipient of several awards, including the Top 50 Influential Papers in Blockchain of 2018, the ACM SIGSOFT Distinguished Paper Award at ICSE2010, the Best Student Paper Award at ICWS2010. He served as BlockSys'19 and CollaborateCom’16 General Co-Chair, SC2'19, ICIOT’18 and IoV’14 PC Co-Chair.
\end{IEEEbiography}
\begin{IEEEbiography}[{\includegraphics [width=1in,height=1.25in] {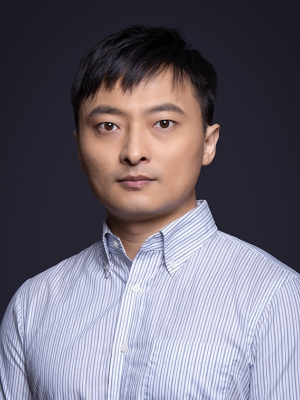}}] {Huawei Huang} \, (Senior Member, IEEE)\  received the Ph.D. degree in Computer Science and Engineering from the University of Aizu, Aizuwakamatsu, Japan in 2016. He is currently an Associate Professor with Sun Yat-Sen University. He was a Research Fellow of JSPS, and an Assistant Professor with Kyoto University, Japan.

His research interests include blockchain and distributed computing. He is currently a Guest Editor for IEEE JOURNAL ON SELECTED AREAS IN COMMUNICA TIONS and IEEE OPEN JOURNAL OF THE COMPUTER SOCIETY, the operation-committee chair for the IEEE Symposium on Blockchain at IEEE SERVICES 2021, and the TPC co-chair of GLOBECOM’2021/ICC’2022 Workshop on scalable, secure, and intelligent Blockchain.
\end{IEEEbiography}

\begin{IEEEbiography}[{\includegraphics[width=1in,height=1.3in]{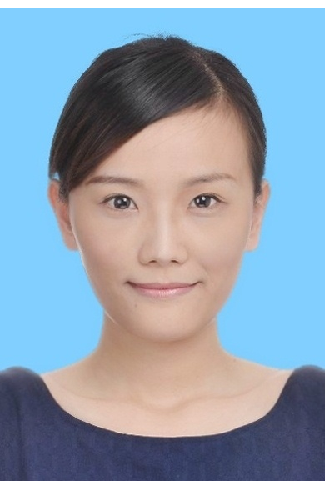}}] {Jiajing Wu}\, (Senior Member, IEEE)\ received the Ph.D. degree from The Hong Kong Polytechnic University, Hong Kong, in 2014. 

In 2015, she joined Sun Yat-sen University, Guangzhou, China, where she is currently an Associate Professor. Her research focus includes blockchain, graph mining, and network science. 

Dr. Wu was awarded the Hong Kong Ph.D. Fellowship Scheme during her Ph.D. study in Hong Kong from 2010 to 2014. She serves as an Associate Editor for IEEE TRANSACTIONS ON CIRCUITS AND SYSTEMS II: EXPRESS BRIEFS.
    
\end{IEEEbiography}
\end{document}